\documentclass{article}

\usepackage{arxiv}

\usepackage{times}
\usepackage{epsfig}
\usepackage{graphicx}
\usepackage{amsmath}
\usepackage{amssymb}
\usepackage{pifont}

\usepackage{booktabs}

\usepackage[utf8]{inputenc} 
\usepackage[T1]{fontenc}    
\usepackage{hyperref}       
\usepackage{url}            
\usepackage{booktabs}       
\usepackage{amsfonts}       
\usepackage{nicefrac}       
\usepackage{microtype}      
\usepackage{cleveref}       
\usepackage{lipsum}         
\usepackage{graphicx}
\usepackage{natbib}
\usepackage{doi}

\title{Dynamic Scene Video Deblurring using Non-Local Attention}

\date{}

\author{ \hspace{1mm} Maitreya Suin
	\And
	\hspace{1mm} A. N. Rajagopalan
}

\renewcommand{\headeright}{}
\renewcommand{\undertitle}{}

\begin{document}
\maketitle

\begin{abstract}
	This paper tackles the challenging problem of video deblurring. Most of the existing works depend on implicit or explicit alignment for temporal information fusion which either increase the computational cost or result in suboptimal performance due to wrong alignment. In this study, we propose a factorized spatio-temporal attention to perform non-local operations across space and time to fully utilize the available information without depending on alignment. It shows superior performance compared to existing fusion techniques while being much efficient. Extensive experiments on multiple datasets demonstrate the superiority of our method.
\end{abstract}


\section{Introduction}
Video deblurring, as a primary problem in the vision and graphics communities, strives to predict latent frames from a blurred sequence.  The camera shake and high-speed motion in dynamic scenes often generate unwanted blur and produce blurry videos. Such videos not only deteriorate the visual quality but also hinder some high-level vision tasks such as tracking \cite{jin2005visual,mei2008modeling}, video stabilization \cite{matsushita2006full}, etc. As more videos are taken using hand-held and onboard video capturing devices, this problem has received great attention in the last decade. The blur in videos is usually a consequence of several interwoven factors like camera shake, object motion, depth variations, etc.

A significant number of works have been proposed \cite{paramanand2011depth,nimisha2018unsupervised,rao2014harnessing,nimisha2018generating,vasu2017local,paramanand2014shape,vijay2013non} where various traditional approaches were adopted for deblurring. Recent works \cite{purohit2019bringing,purohit2020region,mohan2021deep,mohan2019unconstrained,vasu2018non} based on deep convolutional neural networks (CNN) have studied the benefits of replacing the image formation model with a parametric model that can be trained to emulate the non-linear relationship between blurred-sharp image pairs. Unlike single-image deblurring, video deblurring methods can utilize additional information that exists across neighboring frames. Early methods relied on motion compensation of the input frames, either explicitly \cite{su2017deep,mao2016image,hyun2018spatio} or implicitly \cite{zhou2019spatio,zhongefficient}, to aggregate information at a particular location from adjacent frames. \cite{su2017deep,hyun2018spatio,chen2018reblur2deblur} first compute optical flow between a reference frame and neighboring frames and then use the aligned observations to deblur the reference frame. \cite{wang2019edvr} utilizes deformable convolution to align feature maps using learnable offsets. Implicit handling of motion using recurrent networks or 3D convolution has its own drawbacks. 3D convolution \cite{zhang2018adversarial} is computationally heavy and introduces a large number of parameters. For recurrent architectures, the assumption that all previous frames will be automatically aligned and fused in the hidden state remains a problem for frames with large displacement. It is not very easy to extract only the relevant information from a single combined state.

Finding the spatio-temporal relation is critical while fusing information as not all parts of the neighboring frames are equally informative for restoring the current frame due to varying factors such as occlusions, motion, etc. Fusion of incorrect information adversely affects reconstruction performance. We explore the need for non-local operations for spatio-temporal fusion. A non-local self-attention module aims at computing the correlations between all possible pixels within and across frames, which directly resonates with the current goal of spatio-temporal fusion. By nature, such a block does not require any alignment steps. \cite{vaswani2017attention} introduced self-attention based transformer network for natural language processing, \cite{wang2018non} showed a similar non-local approach for classification and recognition. However, extending such approaches to generation tasks is non-trivial. Despite its exceptional non-local processing capabilities, even simpler spatial self-attention can be hard to implement due to its large memory requirement for the image domain. For spatio-temporal operation in videos, it will become significantly more expensive.

In this paper, we present a factorized spatio-temporal self-attention mechanism that contains the essential properties of non-local processing in spatio-temporal domain while being much more efficient. We formulate the entire non-local operation as the composition of three lightweight operators: spatial aggregation, temporal aggregation, and pixelwise adaptive distribution. It requires significantly less memory compared to existing non-local blocks for the same spatio-temporal size while providing superior performance.

To summarize, our contributions are
 \begin{itemize}
 	\item We introduce a factorized spatio-temporal attention as an effective non-local information fusion tool for video deblurring task.
 	\item Extensive experiments and analysis are presented on several video deblurring benchmarks to show state-of-the-art accuracy and interpretability achieved by our architecture.
 	
 \end{itemize}

\section{Related Works}
Early video or multiframe deblurring methods \cite{cho2012video,matsushita2006full} usually assume that there exist sharp contents and interpolate them to help the restoration of latent frames. The main success of these methods is due to the use of sharp contents from adjacent frames. \cite{wulff2014modeling}  develop a novel layered model of scenes in motion and restore latent frames layer by layer. Recently, several end-to-end CNN methods \cite{su2017deep,hyun2017online} have been proposed for video deblurring. To improve the generalization ability, \cite{chen2018reblur2deblur} propose an optical flow based reblurring step to reconstruct the blurry input, which is employed to fine-tune deblurring network via self-supervised learning. \cite{zhang2018adversarial} employ 3D convolutions to help latent frame restoration. \cite{hyun2015generalized} treat optical flow as a line-shaped approximation of blur kernels, which optimize optical flow and blur kernels iteratively. \cite{wieschollek2017learning} recurrently use the features from the previous frame in multiple scales based on a recurrent network. \cite{hyun2017online} develop a spatial-temporal recurrent network with a dynamic temporal blending layer, where they concatenated feature of the current frame and the previous frame and pass through a recurrent network. \cite{zhou2019spatio} fed the previous deblurred frame along with the current blurry frame through their network in a progressive manner and modeled frame alignment and non-uniform blur removal as element-wise filter adaptive convolution processes. \cite{wang2019edvr} develop pyramid, cascading, and deformable convolution to achieve better alignment performance. They have used a simpler temporal and spatial attention strategies. First, they align the neighboring frames, and then at each pixel location, they aggregate the information using convolution.  For spatial attention, they have used simple mask multiplication. In comparison, we resort to more effective non-local processing \cite{buades2005non,wang2018non} using the proposed spatio-temporal attention module where each pixel in the current frame can gather complementary information from all other pixels in all the frames.

\section{Method}
\begin{figure*}[t]
	\centering
	\includegraphics[width = 0.9\textwidth]{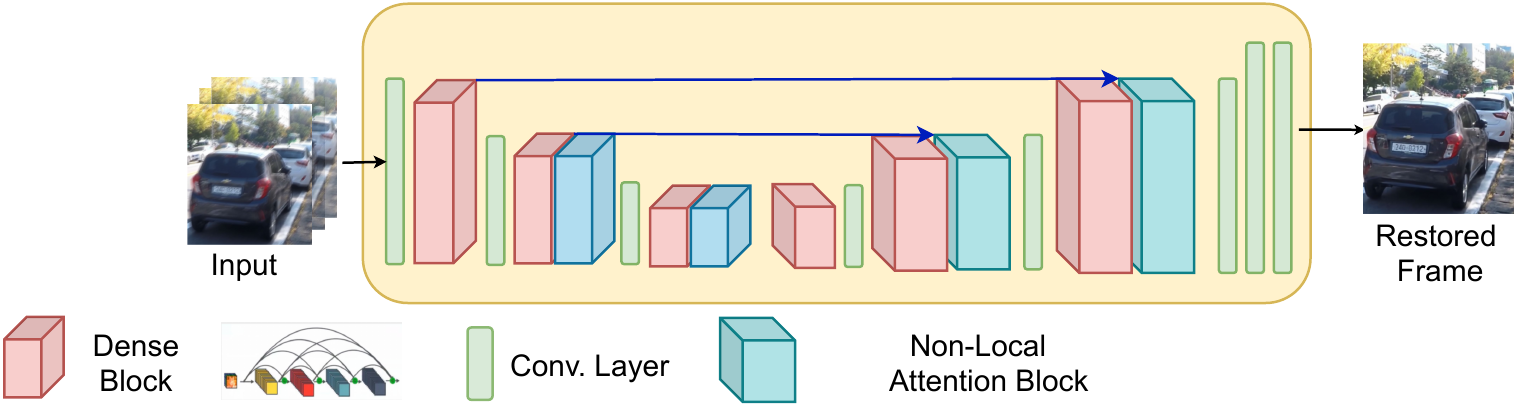}
	\caption{An overview of our method.}
	\label{fig:main_arch}
\end{figure*}

An overview of our network is shown in Fig. \ref{fig:main_arch}. We use a hierchical  encoder-decoder architecture comprising of densely connected modules as the backbone of our restoration network. We use spatio-temporal self-attention blocks to fuse features of the current frame and the neighboring frames.

Non-local means \cite{buades2005non} is a classical filtering algorithm that allows distant pixels to contribute to the filtered response at a location based on patch appearance similarity. This non-local filtering idea was later developed into block-matching algorithm, which was used with neural networks for image denoising \cite{lefkimmiatis2017non}. A similar technique was shown to be successful in the natural language processing domain \cite{vaswani2017attention}. The main building block of \cite{vaswani2017attention} is a self-attention module that computes the response at a position in a sequence (e.g., a sentence) by attending to all positions and taking their weighted average in an embedding space. \cite{wang2018non} proposed a generic non-local operation in deep neural networks to calculate the relation between all possible positions. Given an input feature map of size $T \times H \times W$ (omitting the channel dimension for brevity), the goal of non-local block is to compute the relation $THW \times THW$. But, the tensor of size $THW \times THW$ is huge for videos and to reduce the computational overhead, \cite{wang2018non} typically used $T = 4$, $H = W = 7$. 

For a restoration task like video deblurring, large downsampling will deteriorate pixel-level accuracy. Some works like \cite{ramachandran2019stand} use non-local operations in small blocks inside an image, which hinders its expressibility. Instead, we propose a factorized spatio-temporal self-attention module. Our design is intuitively motivated by examining the flow of information in \cite{hu2018squeeze,cao2019gcnet,li2019expectation} etc, which deploy a squeeze-based aggregation operation in their approach. We gather global information from spatial and temporal domain by performing squeezing operation, and then adaptively distribute it to each pixel of the current frame. We construct three lightweight operations, including spatial squeezing, temporal squeezing, and pixelwise adaptive distribution. For simplicity, we assume batch and channel dimensions to be 1 in the following sections, but it can have any standard values. Given an input tensor $x \in \mathbb{R}^{T \times 1 \times HW}$, we calculate a set of spatial attention maps as
\begin{equation}
	A_s = \text{softmax}_{HW}(f_s(x))
\end{equation}
where $A_s \in \mathbb{R}^{T \times M \times HW}$, $f_s$ is convolutional operation, $\text{softmax}_{HW}$ is softmax along $HW$ and $M$ is the number of attention maps per frame. Next, we elementwise multiply each frame with each of these $M$ attention maps. Let, $A_s^m \in \mathbb{R}^{T \times HW}$ denote the $m^{th}$ attention map. Non-local spatial feature is aggregated for each of the frames using the $m^{th}$ attention map as
\begin{equation}
	\label{eq:sq_spatial}
	G_s^m = \text{S}_{HW}(A_s^m \odot x)
\end{equation}
where $G_s^m \in \mathbb{R}^{T}$, $m \in {1,...,M}$ and $\text{S}_{HW}$ represents the squeeze operation \cite{hu2018squeeze} along $HW$. Now, we  calculate a set of temporal attention maps as
\begin{equation}
	A_t = \text{softmax}_T(f_t(x'))
\end{equation}
where $A_t \in \mathbb{R}^{N \times T}$, $f_t$ is 1D convolutional operator,  $x' \in \mathbb{R}^{1 \times T}$ is the spatially pooled version of the input feature $x$, $\text{softmax}_T$ is softmax operation along $T$. Given the set of temporal attention maps $A_t$ and the spatially aggregated features $G_s$, we apply the $N$ temporal attention maps on each of the $G_s^m: m \in {1,...,M}$ and aggregate temporal information as
\begin{equation}
	G_{st} = G_s A_t^{Tr}
\end{equation}
where $G_{st} \in \mathbb{R}^{MN}$, $G_s \in \mathbb{R}^{M \times T}$, $Tr$ represents transpose operation. Intuitively, each of these $MN$ elements contains global spatio-temporal information, which has been aggregated using the factorized $M$ spatial attention maps and $N$ temporal attention maps resulting in a total of $MN$ possible combinations. After aggregating global information, we adaptively distribute it to each pixel. We generate a pixelwise attention map $A_p$ as
\begin{equation}
	A_p = \text{softmax}_{1}(f_p(x^R))
\end{equation}
where $A_p \in \mathbb{R}^{MN \times HW}$, $f_p$ is 2D convolutional operation, $x^R \in \mathbb{R}^{1 \times HW}$ is the feature map of the current frame. Each pixel will adaptively selectly a particular combination of total $MN$ spatio-temporal attention map using $A_p$. Now, we distribute the global information to each pixel as
\begin{equation}
	y^R = G_{st} A_p
\end{equation}
where $y^R \in \mathbb{R}^{1 \times HW}$ is the output feature map corresponding to the current frame.

\section{Experiments}
\textbf{Implementation Details:} We compare our model with existing works on  GOPRO dataset \cite{nah2017deep} under the standard training and testing settings of previous state-of-the-art methods \cite{pan2020cascaded,nah2019recurrent}. The size of training patch is $256 \times 256$  with minibatch size of 8. We use the ADAM optimizer learning rate of $1e^{-4}$, which decreases to half after every 200 epochs. We implement our algorithm based on the PyTorch on an Ubuntu 16 system, Intel Xeon E5 CPU, and an NVIDIA Titan Xp GPU.

\begin{table*}[t].
	\centering
	\caption{Quantitative evaluations on the GOPRO dataset~\cite{nah2017deep} in terms of PSNR and SSIM.
	}
	\vspace{1mm}
	\label{tab: result-gp-datasets}
	\footnotesize
	\resizebox{0.45\textwidth}{!}{
		\centering
		\begin{tabular}{ccc}
			\hline
			Methods & PSNRs & SSIMs\\
			\hline
			 \cite{tao2018scale}  &30.29 & 0.9014 \\
			 \cite{su2017deep} & 27.31 & 0.8255\\
			 \cite{wieschollek2017learning}& 25.19 & 0.7794\\
			 \cite{hyun2017online} & 26.82 & 0.8245\\
			 \cite{nah2019recurrent} & 29.97 & 0.8947 \\
			 \cite{wang2019edvr} & 26.83 & 0.8426\\
			 \cite{zhou2019spatio} & 28.59 & 0.8608\\
			 Ours & 31.61 & 0.91 \\
	
			\hline
		\end{tabular}
	}
\end{table*}


\begin{figure*}
	\includegraphics[width = 0.95\textwidth]{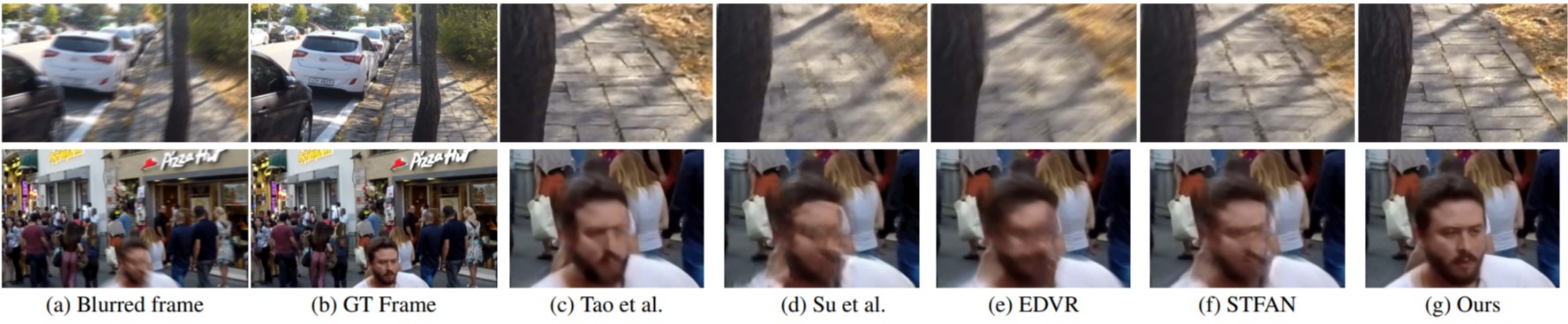}
	\caption{Deblurred results on GOPRO dataset.}
	\label{fig: gp-dataset-vis}
\end{figure*}

\textbf{Quantitative Comparisons:} To evaluate the performance of the proposed algorithm, we compare it against the following state-of-the-art algorithms: \cite{tao2018scale},\cite{su2017deep}, \cite{wieschollek2017learning}, \cite{hyun2017online}, \cite{nah2019recurrent},\cite{wang2019edvr}, \cite{zhou2019spatio}.  Table \ref{tab: result-gp-datasets} shows the quantitative results, where the proposed algorithm performs favorably against the state-of-the-art methods in terms of PSNR and SSIM.
\newline \textbf{Quantitative Comparisons:} Fig. \ref{fig: gp-dataset-vis} show some deblurred results from the testset of \cite{nah2017deep}. We observe that the results of prior works suffer from incomplete deblurring or artifacts. In contrast, our network is able to restore scene details more faithfully, which are noticeable in the regions containing text, edges, etc.

\section{Conclusion}
We have proposed an adaptive approach for video deblurring. The proposed model performs favorably against state-of-the-art methods while being efficient. Such a system can be extended to existing video deblurring methods or other video-processing tasks and will be explored in our future works. Refined and complete version of this work appeared in CVPR 2021.

\bibliographystyle{unsrtnat}
\bibliography{references}  

\begin{thebibliography}{40}
\providecommand{\natexlab}[1]{#1}
\providecommand{\url}[1]{\texttt{#1}}
\expandafter\ifx\csname urlstyle\endcsname\relax
  \providecommand{\doi}[1]{doi: #1}\else
  \providecommand{\doi}{doi: \begingroup \urlstyle{rm}\Url}\fi

\bibitem[Jin et~al.(2005)Jin, Favaro, and Cipolla]{jin2005visual}
Hailin Jin, Paolo Favaro, and Roberto Cipolla.
\newblock Visual tracking in the presence of motion blur.
\newblock In \emph{2005 IEEE Computer Society Conference on Computer Vision and
  Pattern Recognition (CVPR'05)}, volume~2, pages 18--25. IEEE, 2005.

\bibitem[Mei and Reid(2008)]{mei2008modeling}
Christopher Mei and Ian Reid.
\newblock Modeling and generating complex motion blur for real-time tracking.
\newblock In \emph{2008 IEEE Conference on Computer Vision and Pattern
  Recognition}, pages 1--8. IEEE, 2008.

\bibitem[Matsushita et~al.(2006)Matsushita, Ofek, Ge, Tang, and
  Shum]{matsushita2006full}
Yasuyuki Matsushita, Eyal Ofek, Weina Ge, Xiaoou Tang, and Heung-Yeung Shum.
\newblock Full-frame video stabilization with motion inpainting.
\newblock \emph{IEEE Transactions on pattern analysis and Machine
  Intelligence}, 28\penalty0 (7):\penalty0 1150--1163, 2006.

\bibitem[Paramanand and Rajagopalan(2011)]{paramanand2011depth}
Chandramouli Paramanand and Ambasamudram~N Rajagopalan.
\newblock Depth from motion and optical blur with an unscented kalman filter.
\newblock \emph{IEEE Transactions on Image Processing}, 21\penalty0
  (5):\penalty0 2798--2811, 2011.

\bibitem[Nimisha et~al.(2018{\natexlab{a}})Nimisha, Sunil, and
  Rajagopalan]{nimisha2018unsupervised}
Thekke~Madam Nimisha, Kumar Sunil, and AN~Rajagopalan.
\newblock Unsupervised class-specific deblurring.
\newblock In \emph{Proceedings of the European Conference on Computer Vision
  (ECCV)}, pages 353--369, 2018{\natexlab{a}}.

\bibitem[Rao et~al.(2014)Rao, Rajagopalan, and Seetharaman]{rao2014harnessing}
Makkena~Purnachandra Rao, AN~Rajagopalan, and Guna Seetharaman.
\newblock Harnessing motion blur to unveil splicing.
\newblock \emph{IEEE transactions on information forensics and security},
  9\penalty0 (4):\penalty0 583--595, 2014.

\bibitem[Nimisha et~al.(2018{\natexlab{b}})Nimisha, Rajagopalan, and
  Aravind]{nimisha2018generating}
TM~Nimisha, AN~Rajagopalan, and Rangarajan Aravind.
\newblock Generating high quality pan-shots from motion blurred videos.
\newblock \emph{Computer Vision and Image Understanding}, 171:\penalty0 20--33,
  2018{\natexlab{b}}.

\bibitem[Vasu and Rajagopalan(2017)]{vasu2017local}
Subeesh Vasu and AN~Rajagopalan.
\newblock From local to global: Edge profiles to camera motion in blurred
  images.
\newblock In \emph{Proceedings of the IEEE Conference on Computer Vision and
  Pattern Recognition}, pages 4447--4456, 2017.

\bibitem[Paramanand and Rajagopalan(2014)]{paramanand2014shape}
Chandramouli Paramanand and AN~Rajagopalan.
\newblock Shape from sharp and motion-blurred image pair.
\newblock \emph{International journal of computer vision}, 107\penalty0
  (3):\penalty0 272--292, 2014.

\bibitem[Vijay et~al.(2013)Vijay, Paramanand, Rajagopalan, and
  Chellappa]{vijay2013non}
Channarayapatna~Shivaram Vijay, Chandramouli Paramanand, Ambasamudram~Narayanan
  Rajagopalan, and Rama Chellappa.
\newblock Non-uniform deblurring in hdr image reconstruction.
\newblock \emph{IEEE transactions on image processing}, 22\penalty0
  (10):\penalty0 3739--3750, 2013.

\bibitem[Purohit et~al.(2019)Purohit, Shah, and
  Rajagopalan]{purohit2019bringing}
Kuldeep Purohit, Anshul Shah, and AN~Rajagopalan.
\newblock Bringing alive blurred moments.
\newblock In \emph{Proceedings of the IEEE/CVF Conference on Computer Vision
  and Pattern Recognition}, pages 6830--6839, 2019.

\bibitem[Purohit and Rajagopalan(2020)]{purohit2020region}
Kuldeep Purohit and AN~Rajagopalan.
\newblock Region-adaptive dense network for efficient motion deblurring.
\newblock In \emph{Proceedings of the AAAI Conference on Artificial
  Intelligence}, volume~34, pages 11882--11889, 2020.

\bibitem[Mohan et~al.(2021)Mohan, Nithin, and Rajagopalan]{mohan2021deep}
MR~Mahesh Mohan, GK~Nithin, and AN~Rajagopalan.
\newblock Deep dynamic scene deblurring for unconstrained dual-lens cameras.
\newblock \emph{IEEE Transactions on Image Processing}, 30:\penalty0
  4479--4491, 2021.

\bibitem[Mohan et~al.(2019)Mohan, Girish, and
  Rajagopalan]{mohan2019unconstrained}
MR~Mohan, Sharath Girish, and AN~Rajagopalan.
\newblock Unconstrained motion deblurring for dual-lens cameras.
\newblock In \emph{Proceedings of the IEEE/CVF International Conference on
  Computer Vision}, pages 7870--7879, 2019.

\bibitem[Vasu et~al.(2018)Vasu, Maligireddy, and Rajagopalan]{vasu2018non}
Subeesh Vasu, Venkatesh~Reddy Maligireddy, and AN~Rajagopalan.
\newblock Non-blind deblurring: Handling kernel uncertainty with cnns.
\newblock In \emph{Proceedings of the IEEE Conference on Computer Vision and
  Pattern Recognition}, pages 3272--3281, 2018.

\bibitem[Su et~al.(2017)Su, Delbracio, Wang, Sapiro, Heidrich, and
  Wang]{su2017deep}
Shuochen Su, Mauricio Delbracio, Jue Wang, Guillermo Sapiro, Wolfgang Heidrich,
  and Oliver Wang.
\newblock Deep video deblurring for hand-held cameras.
\newblock In \emph{Proceedings of the IEEE Conference on Computer Vision and
  Pattern Recognition}, pages 1279--1288, 2017.

\bibitem[Mao et~al.(2016)Mao, Shen, and Yang]{mao2016image}
Xiaojiao Mao, Chunhua Shen, and Yu-Bin Yang.
\newblock Image restoration using very deep convolutional encoder-decoder
  networks with symmetric skip connections.
\newblock In \emph{Advances in neural information processing systems}, pages
  2802--2810, 2016.

\bibitem[Hyun~Kim et~al.(2018)Hyun~Kim, Sajjadi, Hirsch, and
  Scholkopf]{hyun2018spatio}
Tae Hyun~Kim, Mehdi~SM Sajjadi, Michael Hirsch, and Bernhard Scholkopf.
\newblock Spatio-temporal transformer network for video restoration.
\newblock In \emph{Proceedings of the European Conference on Computer Vision
  (ECCV)}, pages 106--122, 2018.

\bibitem[Zhou et~al.(2019)Zhou, Zhang, Pan, Xie, Zuo, and Ren]{zhou2019spatio}
Shangchen Zhou, Jiawei Zhang, Jinshan Pan, Haozhe Xie, Wangmeng Zuo, and Jimmy
  Ren.
\newblock Spatio-temporal filter adaptive network for video deblurring.
\newblock In \emph{Proceedings of the IEEE International Conference on Computer
  Vision}, pages 2482--2491, 2019.

\bibitem[Zhong et~al.()Zhong, Gao, Zheng, and Zheng]{zhongefficient}
Zhihang Zhong, Ye~Gao, Yinqiang Zheng, and Bo~Zheng.
\newblock Efficient spatio-temporal recurrent neural network for video
  deblurring.

\bibitem[Chen et~al.(2018)Chen, Gu, Gallo, Liu, Veeraraghavan, and
  Kautz]{chen2018reblur2deblur}
Huaijin Chen, Jinwei Gu, Orazio Gallo, Ming-Yu Liu, Ashok Veeraraghavan, and
  Jan Kautz.
\newblock Reblur2deblur: Deblurring videos via self-supervised learning.
\newblock In \emph{2018 IEEE International Conference on Computational
  Photography (ICCP)}, pages 1--9. IEEE, 2018.

\bibitem[Wang et~al.(2019)Wang, Chan, Yu, Dong, and Change~Loy]{wang2019edvr}
Xintao Wang, Kelvin~CK Chan, Ke~Yu, Chao Dong, and Chen Change~Loy.
\newblock Edvr: Video restoration with enhanced deformable convolutional
  networks.
\newblock In \emph{Proceedings of the IEEE Conference on Computer Vision and
  Pattern Recognition Workshops}, pages 0--0, 2019.

\bibitem[Zhang et~al.(2018)Zhang, Luo, Zhong, Ma, Liu, and
  Li]{zhang2018adversarial}
Kaihao Zhang, Wenhan Luo, Yiran Zhong, Lin Ma, Wei Liu, and Hongdong Li.
\newblock Adversarial spatio-temporal learning for video deblurring.
\newblock \emph{IEEE Transactions on Image Processing}, 28\penalty0
  (1):\penalty0 291--301, 2018.

\bibitem[Vaswani et~al.(2017)Vaswani, Shazeer, Parmar, Uszkoreit, Jones, Gomez,
  Kaiser, and Polosukhin]{vaswani2017attention}
Ashish Vaswani, Noam Shazeer, Niki Parmar, Jakob Uszkoreit, Llion Jones,
  Aidan~N Gomez, {\L}ukasz Kaiser, and Illia Polosukhin.
\newblock Attention is all you need.
\newblock In \emph{Advances in neural information processing systems}, pages
  5998--6008, 2017.

\bibitem[Wang et~al.(2018)Wang, Girshick, Gupta, and He]{wang2018non}
Xiaolong Wang, Ross Girshick, Abhinav Gupta, and Kaiming He.
\newblock Non-local neural networks.
\newblock In \emph{Proceedings of the IEEE conference on computer vision and
  pattern recognition}, pages 7794--7803, 2018.

\bibitem[Cho et~al.(2012)Cho, Wang, and Lee]{cho2012video}
Sunghyun Cho, Jue Wang, and Seungyong Lee.
\newblock Video deblurring for hand-held cameras using patch-based synthesis.
\newblock \emph{ACM Transactions on Graphics (TOG)}, 31\penalty0 (4):\penalty0
  1--9, 2012.

\bibitem[Wulff and Black(2014)]{wulff2014modeling}
Jonas Wulff and Michael~Julian Black.
\newblock Modeling blurred video with layers.
\newblock In \emph{European Conference on Computer Vision}, pages 236--252.
  Springer, 2014.

\bibitem[Hyun~Kim et~al.(2017)Hyun~Kim, Mu~Lee, Scholkopf, and
  Hirsch]{hyun2017online}
Tae Hyun~Kim, Kyoung Mu~Lee, Bernhard Scholkopf, and Michael Hirsch.
\newblock Online video deblurring via dynamic temporal blending network.
\newblock In \emph{Proceedings of the IEEE International Conference on Computer
  Vision}, pages 4038--4047, 2017.

\bibitem[Hyun~Kim and Mu~Lee(2015)]{hyun2015generalized}
Tae Hyun~Kim and Kyoung Mu~Lee.
\newblock Generalized video deblurring for dynamic scenes.
\newblock In \emph{Proceedings of the IEEE Conference on Computer Vision and
  Pattern Recognition}, pages 5426--5434, 2015.

\bibitem[Wieschollek et~al.(2017)Wieschollek, Hirsch, Scholkopf, and
  Lensch]{wieschollek2017learning}
Patrick Wieschollek, Michael Hirsch, Bernhard Scholkopf, and Hendrik Lensch.
\newblock Learning blind motion deblurring.
\newblock In \emph{Proceedings of the IEEE International Conference on Computer
  Vision}, pages 231--240, 2017.

\bibitem[Buades et~al.(2005)Buades, Coll, and Morel]{buades2005non}
Antoni Buades, Bartomeu Coll, and J-M Morel.
\newblock A non-local algorithm for image denoising.
\newblock In \emph{2005 IEEE Computer Society Conference on Computer Vision and
  Pattern Recognition (CVPR'05)}, volume~2, pages 60--65. IEEE, 2005.

\bibitem[Lefkimmiatis(2017)]{lefkimmiatis2017non}
Stamatios Lefkimmiatis.
\newblock Non-local color image denoising with convolutional neural networks.
\newblock In \emph{Proceedings of the IEEE Conference on Computer Vision and
  Pattern Recognition}, pages 3587--3596, 2017.

\bibitem[Ramachandran et~al.(2019)Ramachandran, Parmar, Vaswani, Bello,
  Levskaya, and Shlens]{ramachandran2019stand}
Prajit Ramachandran, Niki Parmar, Ashish Vaswani, Irwan Bello, Anselm Levskaya,
  and Jonathon Shlens.
\newblock Stand-alone self-attention in vision models.
\newblock \emph{arXiv preprint arXiv:1906.05909}, 2019.

\bibitem[Hu et~al.(2018)Hu, Shen, and Sun]{hu2018squeeze}
Jie Hu, Li~Shen, and Gang Sun.
\newblock Squeeze-and-excitation networks.
\newblock In \emph{Proceedings of the IEEE conference on computer vision and
  pattern recognition}, pages 7132--7141, 2018.

\bibitem[Cao et~al.(2019)Cao, Xu, Lin, Wei, and Hu]{cao2019gcnet}
Yue Cao, Jiarui Xu, Stephen Lin, Fangyun Wei, and Han Hu.
\newblock Gcnet: Non-local networks meet squeeze-excitation networks and
  beyond.
\newblock In \emph{Proceedings of the IEEE International Conference on Computer
  Vision Workshops}, pages 0--0, 2019.

\bibitem[Li et~al.(2019)Li, Zhong, Wu, Yang, Lin, and Liu]{li2019expectation}
Xia Li, Zhisheng Zhong, Jianlong Wu, Yibo Yang, Zhouchen Lin, and Hong Liu.
\newblock Expectation-maximization attention networks for semantic
  segmentation.
\newblock In \emph{Proceedings of the IEEE International Conference on Computer
  Vision}, pages 9167--9176, 2019.

\bibitem[Nah et~al.(2017)Nah, Hyun~Kim, and Mu~Lee]{nah2017deep}
Seungjun Nah, Tae Hyun~Kim, and Kyoung Mu~Lee.
\newblock Deep multi-scale convolutional neural network for dynamic scene
  deblurring.
\newblock In \emph{Proceedings of the IEEE Conference on Computer Vision and
  Pattern Recognition}, pages 3883--3891, 2017.

\bibitem[Pan et~al.(2020)Pan, Bai, and Tang]{pan2020cascaded}
Jinshan Pan, Haoran Bai, and Jinhui Tang.
\newblock Cascaded deep video deblurring using temporal sharpness prior.
\newblock In \emph{Proceedings of the IEEE/CVF Conference on Computer Vision
  and Pattern Recognition}, pages 3043--3051, 2020.

\bibitem[Nah et~al.(2019)Nah, Son, and Lee]{nah2019recurrent}
Seungjun Nah, Sanghyun Son, and Kyoung~Mu Lee.
\newblock Recurrent neural networks with intra-frame iterations for video
  deblurring.
\newblock In \emph{Proceedings of the IEEE Conference on Computer Vision and
  Pattern Recognition}, pages 8102--8111, 2019.

\bibitem[Tao et~al.(2018)Tao, Gao, Shen, Wang, and Jia]{tao2018scale}
Xin Tao, Hongyun Gao, Xiaoyong Shen, Jue Wang, and Jiaya Jia.
\newblock Scale-recurrent network for deep image deblurring.
\newblock In \emph{Proceedings of the IEEE Conference on Computer Vision and
  Pattern Recognition}, pages 8174--8182, 2018.

\end{thebibliography}

\end{document}